\input{aipcheck}

\documentclass[
    ,final            
  ]
  {aipproc}

\layoutstyle{6x9}
\pdfoutput=1

\begin{document}

\title{3D Numerical Simulations of AGN Outflows in Clusters and Groups}

\classification{98.65.Hb}
\keywords{cooling flow, galaxy clusters,
groups, AGN, jet outflows, ICM, X-rays, 3D simulations}

\author{M. Gaspari}{
  address={Dip. di Astronomia, Universit\`{a} di Bologna, via Ranzani 1, 40127
Bologna, Italy} }

\author{C. Melioli}{
  address={Dip. di Astronomia, Universit\`{a} di Bologna, via Ranzani 1, 40127
Bologna, Italy} }

\author{F. Brighenti}{
  address={Dip. di Astronomia, Universit\`{a} di Bologna, via Ranzani 1, 40127
Bologna, Italy} }

\author{A. D'Ercole}{
  address={INAF - Osservatorio Astronomico di Bologna, via Ranzani 1, 40127
Bologna, Italy} 
}

\begin{abstract}
We compute 3D gasdynamical models of jet outflows from the central
AGN, that carry mass as well as energy to the hot gas in galaxy
clusters and groups. These flows have many attractive attributes
for solving the cooling flow problem: why the hot gas temperature
and density profiles resemble cooling flows but show no spectral
evidence of cooling to low temperatures. Subrelativistic jets,
described by a few parameters, are assumed to be activated when
gas flows toward or cools near a central SMBH. As the jets proceed
out from the center, they entrain more and more ambient gas. Using
approximate models for a rich cluster (A1795), a poor cluster (2A
0336+096) and a group (NGC 5044), we show that mass-carrying jets
with intermediate mechanical efficiencies ($\sim10^{-3}$) can
reduce for many Gyr the global cooling rate to or below the low
values implied by X-spectra, while maintaining $T$ and $\rho$
profiles similar to those observed, at least in clusters. Groups
are much more sensitive to AGN heating and present extreme time
variability in both profiles. Finally, the intermittency of the
feedback generates multiple generations of X-ray cavities similar
to those observed in Perseus cluster and elsewhere. Thus we also
study the formation of buoyant bubbles and weak shocks in the ICM,
along with the injection of metals by SNIa and stellar winds.
\end{abstract}

\maketitle


\section{Introduction}

$\quad$XMM and Chandra spectra of hot gas in galaxy groups and
clusters fail to detect line emission from gas having low
temperatures ($T < T_{vir}/2$), implying the cooling
rate is at least 5-10 times less than previously assumed (\cite{Peterson2006}) 
. The currently most popular explanation for the absence of
detectable cooling is that the hot gas is being heated by the
central AGN (\cite{McNamara2007}). Thus, in order to solve the
cooling flow (CF) problem we compute a large number of
evolutionary models of gas flows, heated mainly by AGN outflows,
in which
gas near the core of cD elliptical galaxies is accelerated in
bipolar subrelativistic "jets", when triggered by a SMBH feedback
mechanism. Both mass and energy are transported out from
the center. 
An inspiration for this work are HI (\cite{Morganti2005}), UV (\cite{Kriss2003}) and 
X-ray (\cite{Risaliti2005}) observations of AGNs, showing
blue-shifted absorption or emission lines along the line of sight.
The outflow velocity is typically several 100-1000 km s$^{-1}$. At
least half of all AGNs exhibit outflows so it is plausible that
they exist in all objects and with substantial covering factors.

The most realistic feedback seems to be linked to the amount of
cooled gas which accretes in the center, so that the total power
injected is $P_{jet} = \epsilon$ $\dot{M}_{cool}$
$c^2$. 
The mechanical efficiency, $\epsilon$, is the free parameter: it
varies in the range $10^{-1} \div 10^{-5}$ and represents all the
unknown physics behind AGN outflows (BH accretion, gas
entrainment, etc.).


\section{Numerical Simulations}

$\quad$ We used a modified version of YGUAZ\'{U}-A, a 3D AMR
(adaptive mesh refinement) code, which solves the hydrodynamic
equations through the "flux vector splitting" of Van Leer (1982),
and is described in \cite{Raga1999}. The simulations start with
the hot gas in hydrostatic equilibrium in the potential well
generated by a NFW dark halo plus the deVaucoleaurs profile of a
cD galaxy. The temperature profile is set to agree with
observations and the density profile is calculated from
hydrostatic equilibrium equation. The finest grids have resolution
of 1-2 kpc, and every model is evolved for 3 Gyr. Radiative
cooling is included assuming a metallicity of 0.3 Z$_{\bigodot}$
(clusters) or 1 Z$_{\bigodot}$ (group). Cooled gas is removed by
adding a mass sink term to the continuity equation:
$-q(T)r/t_{cool}$ (as described by \cite{Brighenti2006}); this
term is used to remove the unphysical clutter of zones containing
cold gas without affecting the hotter flow. The outflow, activated
when gas cools near the SMBH, is generated by imposing momentum
and kinetic energy to the gas located in a small central
cylindrical region (few kpc). We also consider SNIa and stellar
winds in the central elliptical galaxy to estimate metal
enrichment.

The objects chosen for the simulations are rich cluster A1795
(M$_{vir}\sim10^{15}$ M$_{\bigodot}$), poor cluster 2A 0335+096
($\sim2\times10^{14}$ M$_{\bigodot}$) and galaxy group NGC 5044
($\sim4\times10^{13}$ M$_{\bigodot}$).

%



\subsection{Radial Profiles and Cooling Rates}
\paragraph{\textbf{Rich and Poor Cluster}}
$\newline$ In Fig.1 we show that, without heating (solid lines), a
classical CF is soon established in all two systems; the cooling
rate approaches $\sim 300$ M$_\odot$ yr$^{-1}$ after $\sim 1$ Gyr,
which greatly exceeds the observed upper limit. The $\rho$ and $T$
profiles move away from the observed ones. The classical model is
clearly unacceptable. On the other hand, models with heating
illustrate the fundamental characteristic of AGN heated cooling
flows. When the heating is very strong ($\epsilon = 10^{-1}$, long
dash), it can significantly reduce the cooling rate under 1\% of
pure CF, but the temperature in the central regions gets too high
and has the wrong gradient (negative). Conversely, when the
heating is weak ($\epsilon = 10^{-5}$, dot - long dash) , the $T$
profile is less disturbed, but the cooling rate is not reduced by
much ($\sim90\%$). However, when we set an intermediate mechanical
efficiency ($\epsilon = 10^{-3}$, dot - short dash), we are able
to find acceptable models, which means not perfect data overlap,
but a good trend in density and temperature. Indeed, in these good
models with a mean jet velocity of $\sim 5000$ km s$^{-1}$, $\rho$
is not too peaked and $T$ not too shallow, compared to pure CF.
Finally, another interesting feature are the weak shocks at
distant radius (the so-called "ripples", seen by
\cite{Fabian2003}).

\begin{figure}
  \includegraphics[height=.587\textheight,width=\textwidth]{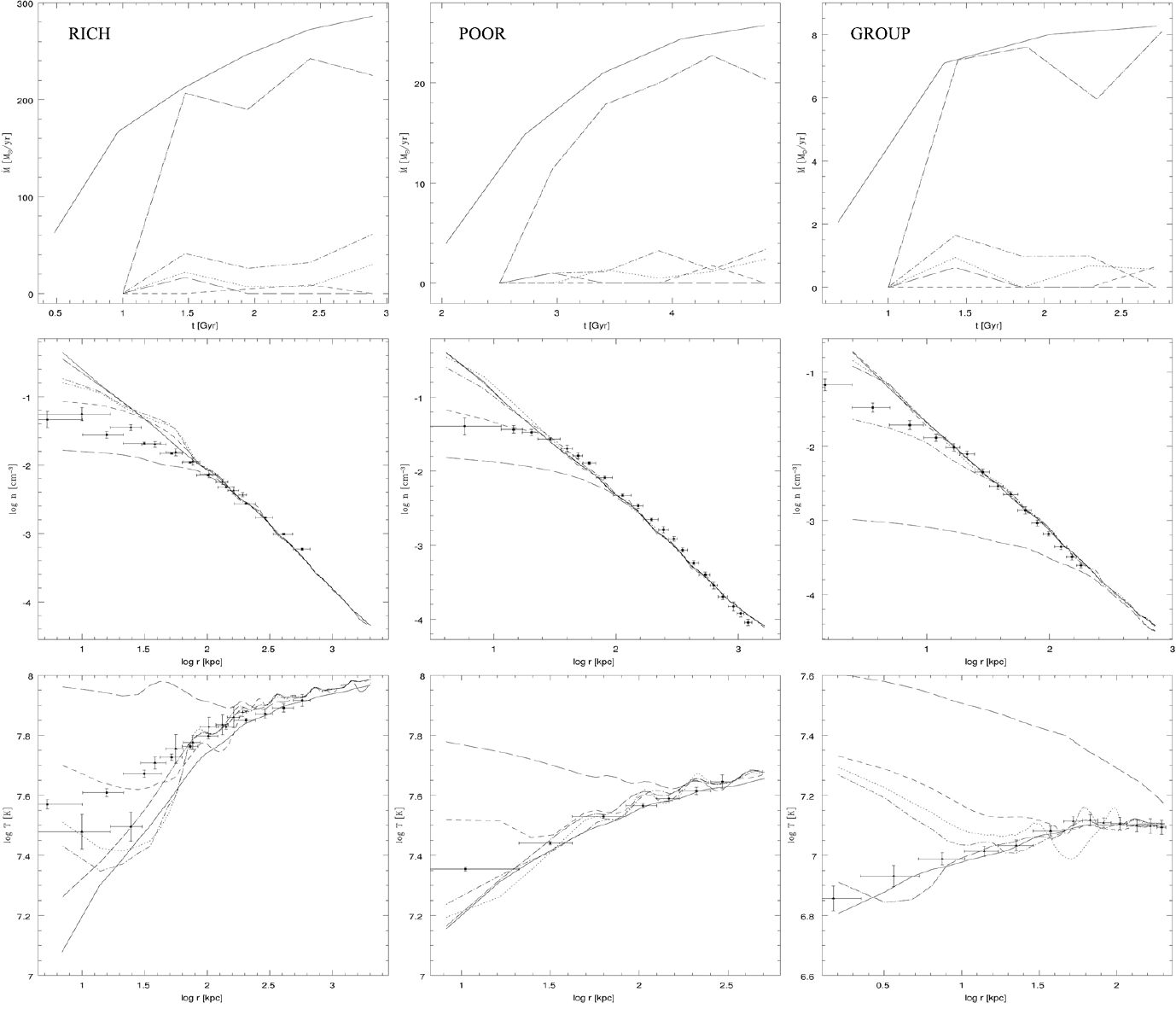}
  \caption{First row: cooling rates as a function of time. Second
  and third rows: radial profiles of $n_e$ and $T$ (emission-weighted)
  after 3 Gyr.
  Solid curves represent pure CF (no heating), while
  others have the following feedback: dot (fixed v$_j = 5\times10^{3}$ km s$^{-1}$),
  short dash (fixed v$_j = 10^4$ km s$^{-1}$), long dash (self-regulated, $\epsilon = 10^{-1}$),
  dot - short dash (self-reg., $\epsilon = 10^{-3}$),
  dot - long dash (self-reg., $\epsilon = 10^{-5}$). Points represent
  observations of A1795, 2A 0335+096 and NGC 5044, from first to third column.}
\end{figure}

\paragraph{\textbf{Group}}
$\newline$ A first analysis reveals that the simulated pure CF has
a $T$ profile very similar to that observed in NGC 5044. That
means the AGN heating does not have to be
so dominant, like in clusters. 
So, when we inject AGN outflows, the heating seems to be always
too little or too much. When $\epsilon = 10^{-5}$ the flow differs
little from the standard CF, except for fluctuations that
accompany the heating episodes. The cooling rate is not
appreciably reduced until $\epsilon = 10^{-3}$, but in this case
$\rho$ and $T$ profiles strongly disagree with observations. Even
with a higher resolution, $\sim 0.5$ kpc, and thus a thinner jet,
the problem still persists. We also noticed that, in groups,
profiles are extremely variable in time, in particular with high
efficiencies, following the AGN feedback activation cycle.

\subsection{Outflow Analysis and 2D Maps}

A detailed analysis of our simulation will be presented elsewhere
(Gaspari et al, in preparation). We give here only a sketch of
the main results.

The main outflows properties of best models are (from group to
rich clusters): velocity $\sim 10^{3-4}$ km s$^{-1}$, with average
power $10^{44-45}$ erg s$^{-1}$. The single jet-event does not
have to be so frequent (every 250 Myr, about 10 Myr long), with
$E_{tot}$ injected a few percent of total "available" BH accretion
energy ($0.1$ $M_{BH}$ $c^2\sim 10^{62}$ erg, with
M$_{BH}\sim10^9$ M$_{\bigodot}$). Outflows with self-regulated
feedback naturally produce several generations of bubbles and weak
shocks, similar to those observed in X-ray (Perseus, M87;
\cite{Fabian2003}), even if it is not present a real relativistic
jet that properly inflates the cavity (Fig.2, left). Bubbles
slowly float upwards buoyantly.
Cavity properties, in best models, are: $5 - 12$ kpc diameter
(from group to rich cluster), with $10^{7} - 10^{8}$ yr lifetime.

In simulations with SNIa and winds, iron maps become asymmetrical
after a few Gyr, showing long filaments. Maximum values are nearly
solar after 3 Gyr (Fig.2, right).

\begin{figure}
  \includegraphics[width=\textwidth]{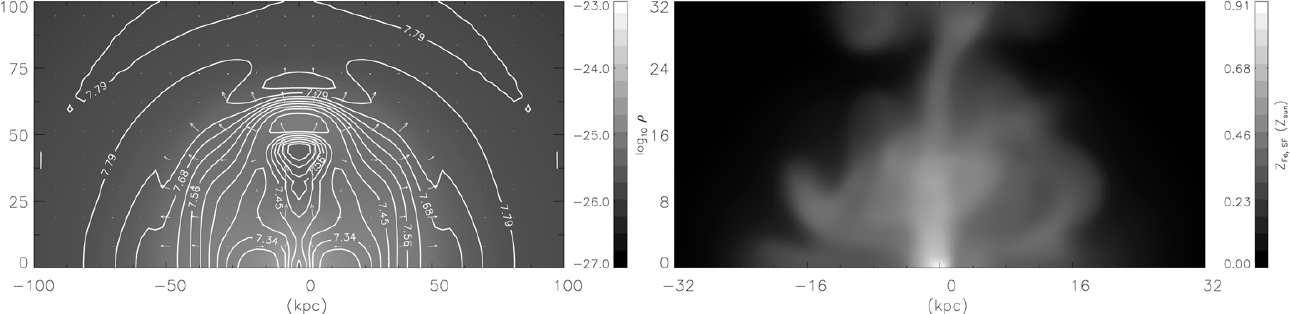}
  \caption{Left: slice through rich cluster center of density,
  with $T$ and $\vec{v}$ superimposed ($\epsilon = 10^{-3}$; t$=2.5$ Gyr).
  Right: Fe-map of the group, emission-weighted along line of sight
  at 3 Gyr ($\epsilon = 5\times10^{-4}$).}
\end{figure}

\section{Conclusions}

$\quad$ Three are the most important features to emphasize. First,
the analysis of the computed gas radial profiles in clusters, rich
and poor, shows that outflows with intermediate efficiencies
($\epsilon = 10^{-3}$) can reproduce the observed $\rho(r)$ and
$T(r)$ with the low cooling rate implied by X-Ray spectra (5 -
10\% of classical CF). Second, models of galaxy groups have a
serious problem: when the heating (AGN plus SNe) is able to stop
or reduce the cooling rate, the $T$ profile strongly disagrees
with the observed one (wrong central gradient, $T$ too high).
Higher computational resolution (thinner jet) helps the density
profile, but the problem still persists. Extreme time variability
of profiles could be explained with the AGN active - quiescent
cycle. Third, intermittency of the feedback naturally generates
multiple X-Ray cavities and weak shocks, similar to those observed
in Perseus and in many other galaxy clusters and groups.

\bibliographystyle{aipproc}   


\begin{thebibliography}{9}






\bibitem{Peterson2006}
J. R. Peterson, A. C. Fabian, 2006, \emph{PR}, 427, 1

\bibitem{McNamara2007}
B. R. McNamara, P. E. J. Nulsen, 2007, \emph{ARA\&A}, 45, 117

\bibitem{Morganti2005}
R. Morganti, C. N. Tadhunter, T. A. Oosterloo, 2005, \emph{A\&A},
444, L9

\bibitem{Kriss2003}
G. Kriss, 2003, \emph{A\&A}, 403, 473

\bibitem{Risaliti2005}
G. Risaliti, S. Bianchi, G. Matt, A. Baldi, M. Elvis, G. Fabbiano,
A. Zezas, 2005, \emph{ApJ}, 630, L129

\bibitem{Raga1999}
A. C. Raga, G. Mellema, S. J. Arthur, L. Binette, P. Ferruit, W.
Steffen, 1999, \emph{RMxAA}, 35, 123

\bibitem{Brighenti2006}
F. Brighenti, W.G. Mathews, 2006, \emph{ApJ}, 643, 120



\bibitem{Fabian2003}
A. C. Fabian, J. S. Sanders, S. W. Allen, C. S. Crawford, K.
Iwasawa, 2003, \emph{MNRAS}, 344, L43









\end{thebibliography}



\end{document}